\documentclass[cameraready]{Interspeech}

\title{Counting Closures in Spanish Trills:\\ A Multi-Corpus Acoustic Study}

\author[affiliation={1},correspondingauthor]{Mateo}{C\'amara}
\author[affiliation={2}]{Maria F.}{Alcala-Durand}

\address{
    $^1$ Signal Processing Applications Group, Information Processing and Telecommunications Center \\
    $^2$ Bioengineering and Optoelectronics \\
    Escuela Técnica Superior de Ingenieros de Telecom., Universidad Polit\'ecnica de Madrid, Spain
}

\email{mateo.camara@upm.es}


\usepackage{amsmath}
\usepackage{tipa}

\begin{document}

\maketitle

\begin{abstract}
The Spanish trill /r/ is canonically described as a short sequence of lingual closures, yet large-scale acoustic evidence across corpora is scarce, and automatic counters locating envelope peaks tend to conflate each closure with its release. We present a closure-based detector that locates closures gated by a quality filter and cross-checked against an independent autocorrelation-based period estimator. Applied to 3,560 well-formed (voiced, periodic) trill tokens from 356 speakers across six Spanish corpora, the detector yields a median of two closures and an inter-closure period near 36ms, matching the descriptive literature on all corpora. At the speaker level, phonotactic context is the only factor with a robust, medium effect: onset trills (word-initial and post-/n,l,s/) show more closures than intervocalic \textit{rr}. We find no robust evidence of a sex effect once closures are counted directly. We report reference values and release a reproducible measurement pipeline for Spanish trills.
\end{abstract}

\section{Introduction}

The Spanish trill /r/ is produced through repeated apical interruptions of the airflow at the alveolar ridge, typically sustained by aerodynamic conditions that enable passive self-oscillation rather than active muscular timing of each contact \cite{sole2002,spajic1996trills}. In Spanish, the trill contrasts phonemically with the tap /\textfishhookr/ in intervocalic position (\textit{perro} vs.\ \textit{pero}) and is required in specific lexical and phonotactic environments, including word-initial \textit{r} and \textit{r} after /n, l, s/ across a syllable boundary \cite{hualde2013sonidos}.

Despite its phonological prominence, quantitative acoustic evidence on trill realization remains fragmented. Traditional descriptions and early acoustic work report a small number of occlusion--release events (typically two to three contacts) and durations on the order of a few tens of milliseconds. However, these estimates are usually derived from limited speaker samples, controlled speech styles, or single varieties \cite{quilis1993,blecua2001}. At the same time, sociophonetic research on Spanish rhotics has largely targeted categorical alternations (e.g., weakening, frication, approximantization) and their conditioning factors, leaving open how fine-grained trill properties such as closure count and closure timing pattern across speakers, contexts, and dialect regions at scale \cite{willis2006,bradley2012}.

This manuscript makes two contributions. The first is methodological: we argue that scalable trill measurement should count \textit{closures} (occlusion events with a verified release) directly rather than peaks of a rectified amplitude envelope. Envelope-peak counters report both the occlusion and the following release as separate events and therefore roughly double the count, in a way that covaries with the spectral density of voicing and hence with speaker characteristics. We describe a closure-based detector that locates band-limited energy minima with verified releases, paired with a fixed, detector-agnostic quality filter and an independent autocorrelation-based period estimator used as an internal consistency check. The second contribution is empirical: applying this pipeline to 3{,}560 trill tokens from 356 speakers across six Spanish corpora, we obtain a median of two closures and an inter-closure period near 36~ms, in line with classic descriptions, and we quantify which speaker- and context-level factors modulate closure structure.

The broader aim is to reconcile the conflicting conclusions of earlier analyses of Spanish trill production by testing whether the prominence attributed to speaker sex and phonotactic context reflects genuine differences in the structure of well-formed trills or the unit chosen to measure oscillatory events. The remainder of the paper reviews related work (Section~2), describes the corpora and token extraction (Section~3), the measurement and statistical pipeline (Section~4), results (Section~5), and robustness analyses (Section~6), and discusses implications (Sections~7--8).

\section{Related Work}

Lingual trilling is a dynamic outcome of articulator mass, stiffness, and a narrow window of aerodynamic conditions. Small changes in pressure or flow can switch a trill to frication or an approximant, which underlies the cross-linguistic and diachronic instability of trills \cite{sole2002,ladefoged1996,spajic1996trills}.

Acoustic descriptions operationalize the trill through occlusion events, segment duration, and voicing, with classic Spanish work providing baseline ranges for closure structure and for the variability of voicing and noise \cite{quilis1993,blecua2001}. A recurring methodological theme is that ``trillness'' is gradient and multi-dimensional, and that estimating closure structure at scale is difficult because trills are frequently partially voiced, irregular, or co-produced with noise.

Sociophonetic studies document substantial dialect- and style-conditioned reduction: Dominican and Jerezano Andalusian Spanish are modally one-closure \cite{willis2006,henriksen2010jerez}, Venezuelan spontaneous speech two-closure \cite{diazcampos2008venezuela}, and Veracruz Spanish can carry the tap/trill contrast through duration when contacts reduce \cite{bradley2012}. Crucially for our study, evidence on speaker sex is mixed: unscripted Peninsular data (median two constrictions, range 0--5) report sex as the strongest predictor \cite{henriksen2023unscripted}, whereas a M\'alaga study found no significant sex effect \cite{zahler2018malaga}. Large-scale, closure-resolved corpus studies remain scarce, requiring reliable phone-level segmentation and scalable measurement \cite{mcauliffe2017}. We address this gap with a closure-anchored detector, effect-size reporting, and robustness checks across datasets.

\section{Data and labeling}
\label{sec:data}
\subsection{Corpora}
We analyze six Spanish speech corpora that provide reliable phone-level segmentation for trill measurement, spanning Peninsular and American varieties and both read and spontaneous speech (Table~\ref{tab:corpora}). All recordings were converted to 16~kHz, mono WAV prior to analysis. Three corpora provide native phoneme-level alignments: DIMEx100~\cite{pineda2004} (millisecond-resolution \texttt{.phn} files), ALBAYZIN~\cite{moreno1993} (phoneme boundaries in SEO transcriptions), and Glissando~\cite{garrido2013} (SAMPA-labeled TextGrids). Three were force-aligned with the Montreal Forced Aligner (MFA)~\cite{mcauliffe2017} using a Spanish acoustic model and pronunciation dictionary: PRESEEA~\cite{preseea2014}, TEDx~\cite{hernandez2018}, and Heroico~\cite{solano2008} (native-speaker subset only). Two further corpora (M-AILABS audiobooks, Common Voice crowd-sourced recordings) were excluded by a pre-specified policy.

\begin{table}[t]
  \caption{The six analyzed corpora. R = read, S = spontaneous. \textbf{Cand.}\ =
  candidate tokens before the quality filter; \textbf{Tok.}\ = the retained
  analysis set (Section~\ref{sec:quality}).}
  \label{tab:corpora}
  \centering
  \footnotesize
  \begin{tabular}{@{}lrrrll@{}}
    \toprule
    \textbf{Dataset} & \textbf{Spk.} & \textbf{Cand.} & \textbf{Tok.} & \textbf{Region} & \textbf{Style} \\
    \midrule
    TEDx        & 138 & 4,144 & 1,481 & Pan-Hisp.\  & S    \\
    Heroico     & 112 & 2,573 & 1,151 & Mex/Sp/Ar   & R    \\
    Glissando   &  27 & 2,226 &   762 & Spain       & R+S  \\
    DIMEx100    &  40 & 6,285 &    81 & Mexico      & R    \\
    ALBAYZIN    &  30 &   346 &    46 & Spain       & R    \\
    PRESEEA     &   9 & 1,262 &    39 & Spain       & S    \\
    \midrule
    \textbf{Total} & \textbf{356} & \textbf{16,836} & \textbf{3,560} & & \\
    \bottomrule
  \end{tabular}
\end{table}

\subsection{Trill token extraction and context labeling}
Tokens were extracted from three contexts where a trill is expected in standard Spanish phonotactics: (i) intervocalic orthographic \textit{rr} (e.g., \textit{perro}, \textit{carro}), (ii) word-initial orthographic \textit{r} (e.g., \textit{rojo}, \textit{rana}), and (iii) orthographic \textit{r} following /n, l, s/ at a syllable boundary (e.g., \textit{enredo}, \textit{alrededor}). Intervocalic single \textit{r} (tap contexts), consonant clusters (e.g., \textit{br, tr, dr, gr, pr, fr}), and coda \textit{r} were excluded to avoid mixing trills with taps and other rhotic realizations. Trill identification combined these orthographic constraints with corpus-specific phone labels, yielding three context categories used consistently across corpora.

\subsection{Quality filter and analysis set}
\label{sec:dataquality}
The orthographic and phone-label criteria yield 16{,}836 aligned candidate tokens across the six corpora. Because automatic alignment and naturalistic recording produce many tokens that are not clean trill realizations (devoiced, fricativized, truncated, or mislabeled), we apply a single fixed, detector-agnostic quality filter as an inclusion criterion (Section~\ref{sec:quality}). A token is analyzed only if its duration is in [50, 200]~ms, at least 80\,\% of its frames are voiced, and its mid-band envelope shows periodic energy modulation in the canonical trill range (periodicity score $\ge 0.40$). The filter retains 3{,}560 of the 16{,}836 candidates (21\,\%) from 356 speakers, with corpus retention ranging from 45\,\% in Heroico to 1.3\,\% in DIMEx100 (Table~\ref{tab:corpora}). It is uniform across corpora; the detector runs only on tokens that pass it, and we test sensitivity to its thresholds in Section~\ref{sec:robustness}. Retention depends on segmentation: the native DIMEx100 \texttt{.phn} boundaries are about half as wide as those of the larger corpora (median candidate 37 vs $\sim$70~ms), so 89\,\% of its candidates fall under the 50~ms floor although 98\,\% pass the voicing rule; PRESEEA (median 30~ms) behaves alike (retention by alignment source and style in the supplement). As a result, no single corpus dominates the filtered sample (TEDx 42\,\%, Heroico 32\,\%, Glissando 21\,\%), unlike envelope-peak analyses, where DIMEx100 supplies three-quarters of the data. The analyzed set is thus clean, voiced, periodic trill realizations: because inclusion requires periodicity, the reference medians characterize well-formed trills by construction, not the rate of trilling in running speech.

\subsection{Speaker metadata}
Speaker sex was taken from corpus metadata where available (ALBAYZIN, Glissando, PRESEEA, TEDx) and from speaker codes for Heroico. DIMEx100 provides no sex metadata, so its speakers are excluded from sex comparisons; no inferred label enters any analysis. Of the 356 speakers, 217 carry a documented sex label (76 F, 141 M). Age, education, and speaking style were harmonized into common bins. We do not analyze country or region: after filtering, each corpus is dominated by one national variety and the cross-national sample is too unbalanced ($\sim$90 speakers vs 3--6) for a defensible dialectal contrast.

\section{Measures and statistical reporting}
\label{sec:measures}

\subsection{Terminology}
\label{sec:terminology}
We use \textit{closure} for the occlusion (contact) event (acoustically a mid-band energy minimum verified by a following release burst) and reserve \textit{cycle} for the full occlusion--release interval. Classic Spanish trill descriptions and most sociophonetic studies count closures annotated manually on spectrograms \cite{quilis1993,blecua2001,henriksen2010jerez,bradley2012,henriksen2023unscripted}, with reports typically ranging from zero to five closures per token. Automatic envelope-peak counters can over-count by reporting both the occlusion and its release as separate maxima; because the number of spurious extra peaks depends on the spectral density of voicing, this bias can covary with speaker characteristics (Section~\ref{sec:results_sex}). We therefore measure \textit{closures} as our primary unit, anchoring detection on energy minima with verified releases rather than on local envelope maxima.

\subsection{Acoustic measures}
For each token, we extract: \textit{closure count} (number of verified closures); \textit{duration} (ms); \textit{inter-closure period} (ms, the mean interval between consecutive closures); \textit{closure rate} (closures per second); and \textit{voicing percentage} (proportion of voiced analysis frames \cite{boersma2001}). Mean $f_0$ is estimated per token by probabilistic-YIN pitch tracking and used to test the counting mechanism (Section~\ref{sec:results_sex}) rather than as an outcome.

\subsection{Closure detection}
\label{sec:detector}
Detection is anchored on energy minima with verified releases. The signal is split into low (60--500~Hz, voicing) and mid (500--3500~Hz, turbulence and release) RMS envelopes \cite{dhananjaya2018trill}. Closure candidates are local minima of the \textit{mid-band} envelope (used instead of a low+mid mean because voicing keeps the low band high and can mask the closure). A candidate is retained if it drops at least 5~dB below and under 40\,\% of the maximum in a 50~ms window, and the combined envelope also drops by at least 3~dB. It is then verified by a release: the maximum of the mid-band envelope 3--25~ms later must reach 40\,\% of the local maximum (one release per closure; unverified candidates are discarded). When a token has at least three closures, a periodicity step removes closures whose adjacent intervals both deviate by more than two standard deviations from the median (with fewer closures it is a no-op). Detection runs on the aligned segment padded by $\pm 20$~ms so that a release just past the boundary can be verified, but only closures inside the original boundaries are kept. Thresholds were set on synthetic trills with known closure counts (16 tests: exact-count recovery, jitter tolerance, vowel rejection, no double counting) and a first pass over the corpora, then frozen; all parameters are identical across corpora, and the two principal thresholds are swept in Section~\ref{sec:robustness}.

\subsection{Quality filter}
\label{sec:quality}
Inclusion is governed by a single fixed rule, applied identically to every corpus (Section~\ref{sec:dataquality}): duration $\in[50,200]$~ms, voicing $\ge 80\,\%$, and an envelope-periodicity score $\ge 0.40$. The periodicity score is the peak of the Pearson-normalized autocorrelation of the mid-band envelope at lags of 25--55~ms (the canonical 18--40~Hz trill range). It depends only on the audio and the segment boundary (not on the closure detector), so the analyzed population is reproducible from the signal alone; it is undefined (set to 0) below 60~ms, so the effective duration floor is 60~ms.

\subsection{Independent cross-detector}
\label{sec:xdetector}
As an internal consistency check, a second estimator counts closures purely from periodicity: it estimates the dominant inter-closure period in 20--60~ms by autocorrelation and reports $n=\mathrm{round}(\text{duration}/\text{period})$. It shares no closure-event logic with the primary detector, but both read the same mid-band modulation, which the filter already requires to be periodic, so per-token agreement within $\pm 1$ closure bounds internal consistency rather than accuracy (Section~\ref{sec:robustness}).

\subsection{Aggregation and inference}
\label{sec:aggregation}
We compare populations defined by speaker metadata and phonotactic context. To avoid pseudo-replication we aggregate to a single observation per speaker for speaker-level factors (sex, age, education, style), and to one observation per speaker$\times$context for the context factor. This yields 356 speaker observations (217 with a documented sex label: 76 F, 141 M). Two-group comparisons (sex) use the Mann--Whitney $U$ test with rank-biserial $r$ as effect size; multi-group comparisons use Kruskal--Wallis $H$ \cite{kruskal1952} with $\varepsilon^2=(H-k+1)/(n-k)$ for $k$ groups, interpreted following Tomczak and Tomczak \cite{tomczak2014} ($<0.01$ negligible, $0.01$--$0.06$ small, $0.06$--$0.14$ medium, $>0.14$ large), and Dunn post-hoc comparisons \cite{dunn1964multiple}. We report 95\,\% bootstrap confidence intervals for group medians (10{,}000 resamples). To control corpus and recording differences jointly, we also fit token-level linear mixed-effects models with a random speaker intercept and fixed effects for context, corpus, duration, and speech style (plus sex on the documented-sex subset), reporting coefficients with 95\,\% CIs (Gaussian, since the count is under-dispersed: variance/mean $\approx 0.3$). $p$-values are Bonferroni-corrected across the test family, and we emphasize effect sizes and confidence intervals over $p$-values alone.

\section{Results}
\label{sec:results}

\subsection{Closure counts match canonical descriptions}
\label{sec:results_canonical}
Across the 3{,}560 analyzed tokens the trill has a median of 2 closures (mean 2.03, SD 0.83) and a median inter-closure period of $\sim$36~ms (corpus medians 30--38~ms), squarely within the two-to-three-contact, 30--50~ms range reported in the descriptive literature \cite{quilis1993,henriksen2010jerez}. Median token duration is 80~ms. These values are stable across all six corpora (Section~\ref{sec:robustness}). Because voicing is part of the inclusion criterion ($\ge 80\,\%$), it is bounded by construction and not analyzed as an outcome.

\subsection{Phonotactic context is the only robust effect}
\label{sec:results_context}
Phonotactic context is the single factor reaching a medium effect, on both closure count ($\varepsilon^2=0.083$) and closure rate ($\varepsilon^2=0.128$). Its direction runs opposite to the intuition that intervocalic \textit{rr} should trill most: intervocalic tokens show the \textit{fewest} closures (mean 1.81) and the \textit{slowest} rate (22.8~Hz), while onset trills show more. Post-/n,l,s/ is fastest (2.21 closures, 27.6~Hz) and word-initial has the most closures and the longest duration (2.30 closures, 91~ms). Part of the count effect tracks duration (word-initial trills are $\sim$12~ms longer), but the rate difference persists at comparable duration: post-/n,l,s/ and intervocalic differ by under 2~ms in duration yet by $\sim$5~Hz in rate. Per-context medians (with 95\,\% bootstrap CIs) all sit near two closures and overlap substantially, so the effect, though robust, is modest. Cells are uneven---intervocalic \textit{rr} 3{,}020/341, post-/n,l,s/ 467/168, word-initial 73/52---so the word-initial estimate is least secure, though its direction holds whichever corpus is left out (Section~\ref{sec:robustness}; full counts in the supplement).

\subsection{No robust evidence of a sex effect}
\label{sec:results_sex}
\begin{figure}[t]
  \centering
  \includegraphics[width=\columnwidth]{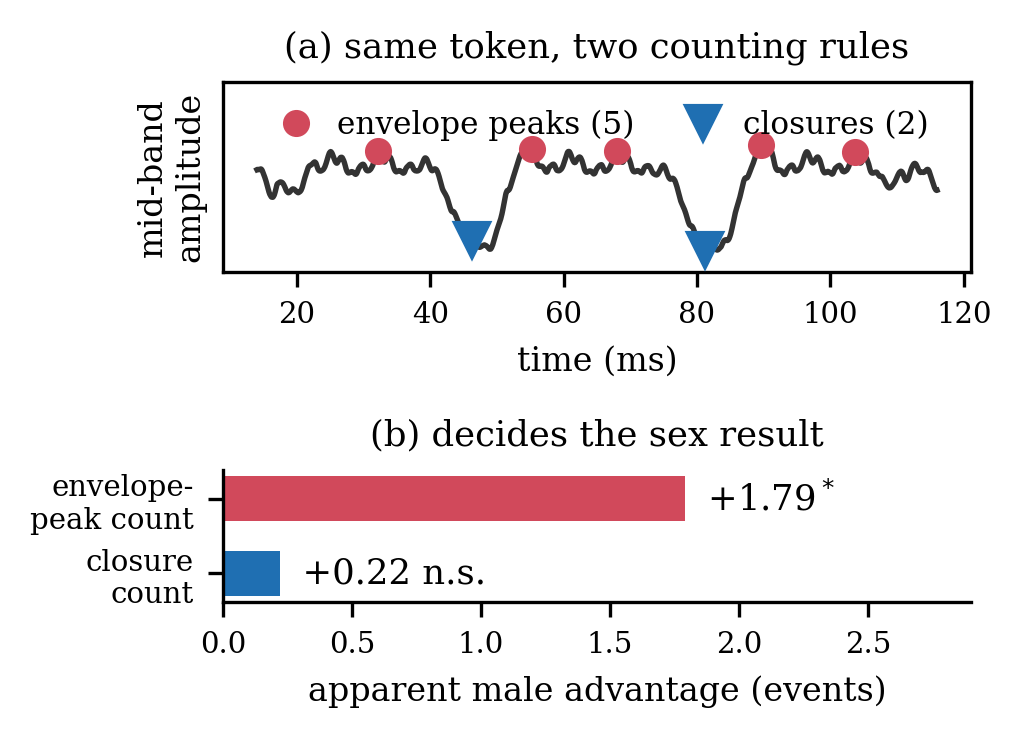}
  \caption{\textbf{The counted unit decides the sociophonetic conclusion.}
  (a)~The mid-band envelope of one trill has two closures (energy
  minima with verified releases) but $\sim$5 envelope peaks. (b)~The
  envelope-peak count inflates an apparent male advantage ($+1.79$ envelope peaks,
  $p<10^{-9}$) that is negligible for closures ($+0.22$, n.s.).}
  \label{fig:summary}
\end{figure}

Counting closures directly, we find no robust evidence of a sex effect on any measure (Figure~\ref{fig:summary}). The median closure count is 2 for both groups. The speaker-level Mann--Whitney test is non-significant ($p=0.20$, rank-biserial $r=0.10$, negligible). The mixed model controlling for context, corpus, duration, and style estimates a male difference of $+0.04$ closures (95\,\% CI $[-0.06, 0.13]$, $p=0.44$; 2{,}417 tokens, 217 speakers). This contrasts sharply with an envelope-peak count of the same audio, whose large effect is a counting artifact: the envelope-peak count covaries with $f_0$ (Spearman $\rho=-0.35$) while the closure count does not ($\rho=+0.01$, $p=0.55$), and the per-token over-count rises as $f_0$ falls ($\rho=-0.37$). Among speakers with independent sex labels, men show $1.79$ more envelope-peak ``cycles'' ($p<10^{-9}$, $r=0.53$) but only $0.22$ more closures (n.s.)---the over-count absorbing most of the gap---suggesting the apparent difference is largely driven by the counting method rather than a robust articulatory difference (Section~\ref{sec:discussion}).

\subsection{Other factors are weak or null}
\label{sec:results_other}
Speaking style has small but significant effects on duration ($\varepsilon^2=0.04$) and closure rate ($\varepsilon^2=0.02$). Read speech is longer and slightly slower than spontaneous speech. Age and education show no significant effect and rest on small, uneven bins, so we report them as null rather than interpret them.

\section{Robustness analyses}
\label{sec:robustness}

\begin{table}[t]
  \caption{Per-corpus literature-anchored checks. Agree.\ = \% of tokens where
  the two detectors fall within $\pm1$ closure (an internal consistency check,
  not ground truth).}
  \label{tab:validation}
  \centering
  \footnotesize
  \begin{tabular}{@{}lrccc@{}}
    \toprule
    \textbf{Corpus} & \textbf{Tok.} & \textbf{Med.\ $n$} & \textbf{Period (ms)} & \textbf{Agree.} \\
    \midrule
    TEDx      & 1,481 & 2.0 & 35.0 & 95.6\,\% \\
    Heroico   & 1,151 & 2.0 & 36.2 & 95.2\,\% \\
    Glissando &   762 & 2.0 & 37.5 & 91.2\,\% \\
    DIMEx100  &    81 & 1.0 & 30.0 & 92.6\,\% \\
    ALBAYZIN  &    46 & 2.0 & 30.0 & 67.4\,\% \\
    PRESEEA   &    39 & 1.0 & 35.0 & 79.5\,\% \\
    \bottomrule
  \end{tabular}
\end{table}

All six corpora fall within the literature-anchored descriptive ranges under the production configuration (Table~\ref{tab:validation}): median closure counts are 1--2 and median periods 30--38~ms. As an internal consistency check, the period cross-detector (Section~\ref{sec:xdetector}) agrees with the closure detector on 91--96\,\% of tokens in the four larger corpora. The two smallest (ALBAYZIN, PRESEEA; 39--46 tokens) are noisier (67--79\,\%) but stay within the descriptive ranges. The counts rest on three convergent references rather than tuning: synthetic trills with known counts, the cross-detector, and the hand-counted literature \cite{quilis1993,henriksen2010jerez}, whose two-closure range the medians reproduce.

The two headline effects are stable across corpora and analysis choices. Leaving out any one corpus, intervocalic \textit{rr} remains the context with the fewest closures and the slowest rate (6/6 hold-outs); the magnitude is small-to-medium ($\varepsilon^2$ $0.05$--$0.13$ for count, $0.08$--$0.16$ for rate, lowest without TEDx). In the mixed model with corpus, duration, and style as covariates, onset contexts keep $+0.22$ closures (post-/n,l,s/, 95\,\% CI $[0.16, 0.28]$) and $+0.23$ (word-initial, $[0.08, 0.38]$) over intervocalic \textit{rr}, and $+2.8$ and $+3.0$~Hz in rate at fixed duration; duration itself adds $0.23$ closures per 10~ms. The period cross-detector reproduces the ranking and counts intervocalic trills \emph{higher} ($2.07$ vs $1.81$), so the effect is not an under-detection artifact. The sex null is likewise uniform across corpora and unchanged in the covariate-adjusted model (Section~\ref{sec:results_sex}).

The conclusions also survive perturbing the two free parts of the pipeline. Across a 15-cell quality-filter sweep (periodicity 0.30--0.50 $\times$ voicing 70--90\,\%), the median stays at two closures, context stays medium ($\varepsilon^2$ $0.06$--$0.09$), and sex stays non-significant throughout ($p\in[0.07,0.65]$). This holds even at the most permissive filter, where the sample is largest, so the null is not a power artifact. A detector sweep over prominence (3, 5, 8~dB) and threshold (0.30--0.50) leaves the outcome unchanged: all six corpora pass for any threshold $\ge 0.40$. The production point (5~dB, 0.40) is thus stable and non-cherry-picked.

Finally, selective attrition could in principle manufacture the null if the filter were sex-biased, as it retains only about a fifth of candidates. It is not: on the 8{,}286 candidates with documented sex (DIMEx100 excluded), retention is essentially equal for women and men (28.1\,\% vs 29.7\,\%; sex\,$\times$\,inclusion $\chi^2$ $p=0.14$; speaker-level $p=0.36$), and male candidates are not less voiced (speaker-level $p=0.64$), so the voicing rule does not remove male tokens preferentially (per-sex table in the supplement).

\section{Discussion}
\label{sec:discussion}

At scale, the Spanish trill centers on two closures and a $\sim$36~ms period, consistent with classic and corpus descriptions \cite{quilis1993,henriksen2010jerez,diazcampos2008venezuela}. These reference values describe well-formed, voiced, periodic trills: the filter removes the reduced variants (fricated, approximantized, devoiced) that are frequent in casual speech, so they characterize the target configuration and say nothing about how often trilling succeeds in running conversation, a question that requires modeling realization before closure structure.
The context effect is partly duration-driven but persists at fixed duration (Section~\ref{sec:robustness}) and is not a detection artifact: the independent period cross-detector, insensitive to closure depth, reproduces the ranking and, if anything, counts intervocalic trills \emph{higher}. We read it as robust but modest, compatible with aerodynamic accounts of position-sensitive trilling \cite{sole2002}.

Our sharpest finding is methodological: an envelope-peak counter produces a large, significant sex effect on the same audio while a closure-based counter does not. Envelope-peak counting registers a maximum for both the occlusion and its release, so the count scales with $f_0$ and harmonic density (over-count vs.\ $f_0$: $\rho=-0.37$; Section~\ref{sec:results_sex}), themselves correlated with sex. Anchoring on energy minima with verified releases removes this confound. The resulting null aligns with the M\'alaga study of Zahler \cite{zahler2018malaga} and tempers reports of sex as the strongest predictor of trill ``cycles'' \cite{henriksen2023unscripted}. A remaining limitation is the absence of a manually annotated benchmark, which the cross-detector cannot replace since both read the same envelope: future work should compare the detector against expert closure counts on a balanced subset, to be released with the pipeline (Section~\ref{sec:code}).

\section{Conclusion}

Across six Spanish corpora, well-formed voiced trills center on two closures with a $\sim$36~ms period; phonotactic context is the only robust effect. And, contrary to the intuition that the ``strong'' intervocalic \textit{rr} should trill most, it is the onset contexts (word-initial, post-/n,l,s/) that show the most closures and intervocalic trills the fewest, although the word-initial estimate (73 tokens) is tentative pending a balanced corpus. Still, we find no robust evidence of one for speaker sex once closures are counted directly. The methodological message is that the unit of count (closures versus envelope peaks) can determine whether a sociophonetic effect appears: because envelope-peak counts covary with $f_0$, earlier reports attributing trill differences to speaker sex or other demographic factors may partly reflect the counting method and warrant re-examination.

\clearpage
\section{Code availability}
\label{sec:code}

The measurement pipeline, statistical scripts, and supplementary material are available at \url{https://github.com/MateoCamara/trillscope}.

\section{Generative AI Use Disclosure}
The authors used generative AI for editing and polishing the manuscript.

\section{Acknowledgements}
The authors would like to thank Stefanie Shattuck-Hufnagel for identifying this research line, and José Luis Blanco and Juan Ignacio Godino for the interesting discussions.

\bibliographystyle{IEEEtran}
\bibliography{mybib}

\end{document}